\documentclass[12pt]{article}
\usepackage{graphicx}
\def\lvbr{\vphantom{\frac{\Big[\Big]}{\Big[\Big]} }}% large vertical
\def\svbr{\vphantom{\frac{E^E_E}{E^E_E} }}% small vertical brace for tables

\title{A New Constraint on Strongly Coupled Field Theories}

\author{Thomas Appelquist$^a$, Andrew G. Cohen$^b$, and Martin
  Schmaltz$^c$\ \thanks{\tt thomas.appelquist@yale.edu,
    cohen@bu.edu, schmaltz@slac.stanford.edu}\\ \\
  \small \sl $^a$Department of Physics, Yale University, New Haven, CT 06511\\ \\
  \small \sl $^b$Department of Physics, Boston University, Boston, MA
  02215 \\ \\
  \small \sl $^c$SLAC, Stanford University, Stanford, CA 94309\\ \\
  }

\begin{document} 
\setlength{\baselineskip}{24pt}
\begin{titlepage}
\maketitle
\begin{picture}(0,0)(0,0)
\put(295,370){BUHEP-99-2}
\put(295,360){YCTP-P2-99}
\put(295,350){SLAC-PUB-8045}
\end{picture}
\vspace{-36pt}

%\date{January, 1999}

\begin{abstract}
We propose a new constraint on the structure of strongly coupled,
asymptotically free field theories. The constraint takes the form of an
inequality limiting the number of degrees of freedom in the
infrared description of a theory relative to the number of
underlying, ultraviolet degrees of freedom. 
We apply the inequality to a variety of
theories (both supersymmetric and nonsupersymmetric), where it
agrees with all known results and leads to interesting new
constraints on low energy spectra. 
We discuss the relation of this constraint to Renormalization
Group $c$-theorems.
\end{abstract}
\thispagestyle{empty}
\setcounter{page}{0}
\end{titlepage}

\section{Introduction}

Four dimensional field theories have been remarkably successful at
describing nature at energies less than several hundred GeV.
Unfortunately progress at higher energies has been frustrated by a
dearth of general theoretical tools that apply to strongly coupled
models. Our understanding of field theory comes largely from
perturbation theory (which applies to weakly coupled systems) and
from QCD (where specific strong dynamics may be compared to
experiment). There are many examples where this understanding is
inadequate---for example, even the question of chiral symmetry
breaking in a QCD-like gauge theory with a large number of flavors
is unsettled
\cite{Appelquist:1996dq,Appelquist:1998rb,Chen:1997jj,Mawhinney:1997jm,Shuryak:1997as}.

Recently new tools have appeared in the context of supersymmetric
gauge theories. Known collectively as ``duality'', these ideas have
produced convincing pictures of the pattern of symmetry breaking in
many strongly coupled supersymmetric theories. The wide variety of
low-energy phenomena that appear is remarkable, including dual
gauge groups, conformal fixed points, chiral symmetry breaking,
{\it etc}. These results are obtained without a detailed solution
for the dynamics at strong coupling, but rely on symmetries,
inspired guesswork, and general properties of supersymmetry.
This shows that general constraints on
the low energy properties of strongly coupled field theories are
enormously  useful, especially when a complete solution is
unavailable.

The most powerful general constraint known is the anomaly matching condition
introduced by 't Hooft \cite{'tHooft:1979bhprime}.  Generally, we may define an
anomaly as a residue of the pole in a particular multi-current
correlation function . 
As discussed by 't Hooft, this number is independent of
renormalization scale, and may therefore be computed at short
distances, or equally well at long distances:
\begin{equation}
  \label{eq:anomaly}
  {\cal A}_{IR}={\cal A}_{UV} \ .
\end{equation}
As the residue of a pole, the anomaly only receives contributions 
from physical massless degrees of freedom. If the short distance
theory is weakly coupled (like an asymptotically free gauge theory)
or calculable by other means, the anomaly condition provides an
immediate relation of the massless spectrum to the short distance
physics, constraining the appearance of massless fermions and
Nambu-Goldstone bosons. Anomaly matching, as implementation of this
condition is often called, has led to useful constraints on the
possible low energy realizations of chiral gauge
theories \cite{Peskin:1982muprime,Eichten:1986fs}, as well as QCD-like
(vector-like) gauge
theories \cite{Coleman:1980mx,Vafa:1984tf}. Anomalies have also played
a fundamental role in 
discovering and checking the dualities of supersymmetric gauge
theories \cite[and references therein]{Intriligator:1996au}.

In this paper we propose a new constraint on the structure of
strongly coupled field theories. Before stating and discussing this
constraint, we note that, although the anomaly condition does not
forbid the appearance of additional (vector-like) massless particles,
it is 
usually assumed that the spectrum contains no massless particles
that this condition does not require.
That is, if there are relevant operators not forbidden by
symmetries that would produce masses, it is (technically) unnatural
to assume that these operators are absent. Consequently we might
say that nature generally abhors massless particles.

In fact when faced with the task of guessing the massless spectrum
of a strongly coupled field theory, we are often guided by the idea
that the number of  massless particles is as small as possible.
Since the anomaly condition can always be satisfied by a massless
spectrum identical to the ultraviolet degrees of freedom, this
would  disfavor massless composites if the number of such
composites is too large.

In elevating this casual notion to a formal principle, we need 
precise definitions of the number of degrees of freedom in both the
infrared and ultraviolet.  Although there are no unique such objects,
we will choose to define quantities related to the free energy of the
field theory. We will consider only renormalizable theories,
for which the free energy may be rendered finite and
cutoff independent by adjusting the vacuum energy to zero,
renormalizing a finite set of parameters, and then removing the cutoff
(holding physical quantities and the temperature fixed). For reasons
described later, we will consider only asymptotically free theories.

In terms of this properly renormalized free energy per unit volume,
$\cal F$, (which is also equal to minus the pressure), the quantity
that we will use to characterize the number of infrared degrees of freedom is
\begin{equation}
  \label{eq:firdef}
  f_{IR} \equiv - \lim_{T\to 0} \frac{\cal F}{T^4} \frac{90}{\pi^2}
\end{equation}
where $T$ is the temperature. For a free field theory,
$f_{IR}$ is simply the number of massless bosons plus $7/8$ times
the number of massless fermions. For an asymptotically free theory,
the corresponding expression in the large $T$ limit mesaures the
ultraviolet degrees of freedom in a similar way:
\begin{equation}
  \label{eq:fuvdef}
  f_{UV} \equiv - \lim_{T\to \infty} \frac{\cal F}{T^4} \frac{90}{\pi^2}.
\end{equation}

Our qualitative discussion above suggests the new constraint $f_{IR}
\le f_{UV}$. In Section 2 we formulate this idea precisely and
describe how this inequality (assuming that it is correct) leads to
restrictions on the physical properties of strongly coupled field theories.
Two examples are
considered: a supersymmetric $SU(N)$ gauge theory with $F$
flavors, and a non-supersymmetric version of the same theory.  In
both cases the inequality will constrain the low energy structure.
In Section 3 we describe a (failed) route to a
proof of the inequality. The line of argument is nevertheless
interesting, and 
leads to a deeper understanding of the inequality and its relation
to so-called ``$c$-theorems'' \cite{Zamolodchikov:1986gt}.
In Section 4, we discuss the $T$ dependence for the two examples
mentioned above. In Section 5 we apply the inequality to a variety
of strongly coupled field theories. Finally, in Section 6 we
summarize and conclude.

\section{The Inequality}

Our conjectured inequality is
\begin{equation}
  \label{eq:ineq}
  f_{IR} \le f_{UV} \
\end{equation}
%subject to the condition that the limits in
%Eqs.~(\ref{eq:firdef},\ref{eq:fuvdef}) exist.  
These limits are well
defined for theories with  both  UV and IR fixed points.  However
 the inequality can be
violated in the presence of non-trivial UV fixed points, as we show in 
section 5.2. Hence our
examples will involve asymptotically free gauge theories, and most
will be infrared free as well, although the IR degrees of freedom may
be different from those in the UV.

In field theories with weakly coupled  fixed points the free
energy, appearing in the definition of $f$,  may be computed
perturbatively. To zeroth order in couplings
the low-temperature free energy density in three spatial dimensions is
\begin{equation}
{\cal F}_{free}(T) \simeq -{\pi^{2} T^4\over 90} [ N_B + {7
  \over 8} 2 N_F],
\label{DofF}
\end{equation}
where $N_B$ is the number of massless (real) bosonic fields, and
$N_F$ is the number of massless (two-component) fermionic fields.
We have neglected the contributions of any massive fields, which
vanish exponentially as $T\to 0$. A similar expression applies in
the infinite $T$ limit, with $N_B$ and $N_F$ including massive as
well as massless fields. These expressions are exact in the case of
free fixed points and approximately correct for
theories governed by weak fixed points. For
specific theories we may include perturbative corrections.

\subsection{SUSY Example}

For our first example we consider a SUSY $SU(N)$ gauge theory with
$F$ flavors (``quarks'' and ``antiquarks'') of massless fermions
and associated superpartners.
The theory has a free UV fixed point if the number of flavors
is less than $3$ times the number of colors, $F < 3N$. In this
case the quantity $f_{UV}$ may be calculated using Eq. (\ref{DofF})
to give
\begin{equation}
  \label{eq:SQCDUV}
  f_{UV} = \left[2(N^2-1)+ 4 N F\right](1+\frac{7}{8})\ .
\end{equation}

The analysis of Seiberg \cite{Seiberg:1995pq} suggests that the infrared
behavior of this theory is alternatively described through the use of $F$
flavors of massless magnetic quarks transforming according to the
fundamental representation of a dual gauge group $SU(F-N)$, along
with $F^2$ massless ``meson'' chiral superfields. This
theory is infrared free provided $F \le 3N/2$. Under these
circumstances $f_{IR}$ is:
\begin{equation}
  \label{eq:SQCDIR}
  f_{IR} = \left[2((F-N)^2-1)+ 4 (F-N) F+2
    F^2\right](1+\frac{7}{8})\ .
\end{equation}
Thus our fundamental inequality becomes
\begin{equation}
  \label{eq:sineq}
  2[(F-N)^2-1]+ 4 (F-N) F+2
    F^2 \le 2(N^2-1)+ 4 N F
\end{equation}

Because $f_{IR}$ grows quadratically with the number of flavors, this
inequality 
limits the values of $F$ for which the low energy theory can
consist of massless magnetic degrees of freedom with infrared free
coupling. Remarkably, this inequality gives the bound $F \le (3/2)N$,
corresponding precisely to the boundary of the weak magnetic phase
determined by the analysis of Seiberg \cite{Seiberg:1995pq}. At the boundary
$F=(3/2)N$ the inequality is saturated\footnote{The simplest example
is the case $N=2$: $SU(2)$ gauge theory with 3 flavors. The theory
confines and has an infrared-free dual description containing only
a ``meson'' superfield, and $f_{UV}=f_{IR}=30 (1+ 7/8)$.}.
We will show that the inequality continues to hold
for $F>(3/2)N$ in Section 5.2.

\subsection{Non-SUSY Example}

For our second example we consider the non-supersymmetric version
of the same $SU(N)$ gauge theory, with $F$ massless quarks (and
antiquarks). The theory has a free UV fixed
point for $F < 11N/2$. Based on real QCD we expect the $SU(F)\times
SU(F)$ chiral symmetries of this theory to be realized in the
Nambu-Goldstone mode---at least for small enough $F/N$. If we
assume that this is the case, the IR theory consists of $F^2-1$
Nambu-Goldstone bosons. The derivative interactions of these
particles are irrelevant in the infrared, and consequently this
theory is described by a free IR fixed point.

At these free UV and IR fixed points we may use Eq. (\ref{DofF}) to
compute $f_{IR}$ and $f_{UV}$:
\begin{eqnarray}
  \label{eq:QCD}
  f_{IR} = F^2 -1 \nonumber\\
  f_{UV} = 2(N^2-1) +\frac{7}{8} 4NF\ ,
\end{eqnarray}
and our inequality becomes
\begin{equation}
  \label{eq:QCDineq}
 F^2 -1 \le2(N^2-1) +\frac{7}{8} 4NF\ ,
\end{equation}
or, since $F$ must be positive
\begin{equation}
  \label{eq:QCDconstraint}
 F \le 4 \sqrt{N^2 - \frac{16}{81}}\ .
\end{equation}
Since $N$ must be 2 or larger, and $F$ and $N$ must both be
integral, this is equivalent to $F < 4N$. Remarkably, our
inequality says that for the number of flavors larger than or equal
to four times the number of colors, this gauge theory cannot break
the full set of chiral symmetries!

This new bound on the onset of the chiral phase transition ($F \leq
12$ for $SU(3)$) is well above the transitional values suggested by
preliminary lattice simulations \cite{Chen:1997jj,Mawhinney:1997jm}. It
is very close,  
however, to the value that emerges from the use of a continuum gap
equation together with the assumption that the coupling is governed by
an infrared fixed point appearing in the perturbative $\beta$
function \cite{Appelquist:1998rb}. In fact, a combination of the
ladder gap equation 
and the two-loop beta function give a critical value $F^{crit}/N =
(100N^2 - 66 ) / (25N^2 - 15 ) (\rightarrow 4$ as $ N \rightarrow
\infty$). The reliability of this result is far from clear, however,
since higher order effects are not obviously small.  So whether the
chiral phase transition saturates the inequality in this way or
takes place at a lower value of $F/N$ remains an open question.

\section{Relation to $c$}
\label{sec:c-theorem}

Having shown that the inequality Eq. (\ref{eq:ineq}) is consistent with
other analyses of the SUSY $SU(N)$ theory and that it leads to a new 
result for QCD-like theories, we next discuss why it might be 
true generally.
As an attempt at proof we may define a function
$f(T)$ at all scales in an obvious way, as minus the free energy
density divided by $T$ raised to the number of spatial dimensions
plus one (this extension away from 4 space-time dimensions will
prove useful shortly):
\begin{equation}
  \label{eq:f}
  f(T) \equiv -\frac{\cal F}{T^{d+1}} \Omega_d
\end{equation}
where $\Omega_d$ is a constant chosen such that the contribution to
$f(T)$ from a free bosonic degree of freedom is 1. The quantities
$f_{IR}$ and $f_{UV}$  are just the limits of this function as $T$
approaches zero and $\infty$ respectively.

As a first step we differentiate the function
$f$ with respect to $T$. Using the standard relations $T\partial {\cal
  F}/\partial T = -u-p$, ${\cal F} = -p$ where $p$ is the pressure and
$u$ is the internal energy density, we have
\begin{equation}
  \label{eq:trace}
  T\frac{\partial f}{\partial T} = \Omega_d \frac{u-d p}{T^{d+1}} \equiv
  \Omega_d \frac{\theta}{T^{d+1}}
\end{equation}
where $\theta$ is the (thermal average of the) trace of the
energy-momentum tensor. For a conformally invariant theory, the
trace of the energy-momentum tensor is zero. Under these
circumstances we see that $f$ is a constant, and $f_{IR}$ is equal
to $f_{UV}$. Of course the theories that we are interested in are
not conformally invariant---the lack of conformal invariance arises
from a scale dependence of coupling constants through
renormalization. Consequently we expect the difference between
$f_{UV}$ and $f_{IR}$ to arise from the renormalization group flow
from the ultraviolet to the infrared. If $\theta$ is positive along
this trajectory, $f_{IR}$ will necessarily be smaller than
$f_{UV}$, proving our inequality.

Thus our inequality would follow from a positivity condition on the
thermal average of the trace of the energy-momentum tensor; that is,
from positivity of $u-d p$. For non-interacting systems, massive
modes always have $p < u/d$ whereas massless modes have $p =
u/d$. Even for interacting classical systems we expect these
conditions to remain valid. Unfortunately the situation in quantum
theories is not so simple \cite{CastroNeto:1993ie,Sachdev:1993pr}.

Consider, for example, a classically scale-invariant gauge field
theory. In this case the trace of the energy-momentum tensor is
given by:
\begin{equation}
  \label{eq:emtrace}
  \theta = 2\frac{\beta}{g} \hbox{Tr} G^2
\end{equation}
where $\hbox{Tr} G^2$ is the trace over gauge indices of the square of
the gauge field-strength, as well as over the thermal density matrix,
and $\beta$ is the RG beta function.
Note that, at least in perturbation theory, the thermal average of the
field-strength squared is negative (magnetic fluctuations are
less-well screened than electric fluctuations). Therefore a negative
beta function leads to a positive $\theta$, as we
desire.

Unfortunately this observation immediately suggests examples of
{\em negative} $\theta$. If the low energy theory is a gauge field
theory governed by a free infrared fixed point, then the $\beta$
function will be positive at weak coupling where $\hbox{Tr} G^2$ is
known to be negative. This is realized if the low energy theory is
either an abelian theory with massless fermions or a non-abelian
theory with matter content sufficient to render it infrared free.
The SUSY $SU(N)$ theory in the weak magnetic phase, discussed in
the previous section, is precisely such an example.

Of course this is not a counter-example to our conjectured
inequality: the fact that $f(T)$ is not monotonic does not
contradict the inequality involving $f_{IR}$ and $f_{UV}$. ( We
have already noted that the SUSY $SU(N)$ theory in the weak
magnetic phase does in fact satisfy the inequality.) It means,
however, that a proof of this inequality will be more involved than
the simple argument used here. 

%In addition to constraining the infrared behavior of field theories,
%the inequality can also restrict the general form of renormalization
%group flows. If we could construct a theory for which $\theta$ is
%always negative, the inequality would be violated. An example of such
%a theory might be a gauge theory  in which $\theta$ is given by
%Eq. (\ref{eq:emtrace}), $\beta$ is always positive, and the coupling
%is weak so that $\hbox{Tr} G^2$ is always negative. Of course we
%continue to require a UV and an IR fixed point.
%We know of no example of a four dimensional gauge theory of this kind.

This discussion also indicates why we restrict our attention to
asymptotically free theories. Negative contributions to $\theta$ 
decrease $f_{UV}-f_{IR}$; if these contributions persist over a large
temperature range, the inequality will be violated.
Since operators in the Hamiltonian contribute to $\theta$ according to 
their scaling dimension,  positive operators with a coupling constant
of negative mass dimension (positive ``irrelevant'' operators)
make a {\em negative} contribution to $\theta$. 
A renormalizable theory with a non-trivial UV fixed point may have such an
operator which can make a negative contribution to $\theta$ over a
large range of temperature, invalidating the inequality. (An explicit
example of this type is mentioned at the end of section 5.2.) We thus
consider only asymptotically free theories.

Note that had the function $f(T)$ been monotonic, we would have
proven a ``$c$-theorem'': the existence of a function that is
monotonic along RG trajectories. For example, in one spatial
dimension the function $f(T)$ {\em is} monotonic, since the energy
density is always greater than or equal to the pressure in the
thermal state. But the value of this function at any fixed point
(where $\theta = 0$) is simply the conventionally defined central
charge of the corresponding conformal field
theory\footnote{Although our argument
  involves a flow in temperature, the usual RG arguments and
  dimensional analysis may be used to rewrite everything in terms of a
  change in scale parameter, $\mu$.}. The decrease in central charge
between fixed points along an RG trajectory is the $c$-theorem of
Zamolodchikov \cite{Zamolodchikov:1986gt}.

%Our analysis implies that the existence of a corresponding $c$-theorem 
%(with the same $c$ charge) in four
%space-time dimensions would lead to the inequality Eq.
%(\ref{eq:ineq}). If such a theorem does exist, the $c$-function is
%clearly {\em not} the function $f(T)$. Any other monotonic function
%that is equal to $f(T)$ at conformal fixed points is equally
%suitable for proving our conjecture. We have been unable to find
%such a function. Neither have we been able to establish the
%impossibility of such a function. Note that the existence of such a
%$c$-theorem, while providing a proof of Eq. (\ref{eq:ineq}), is not a
%necessary condition for the correctness of our
%much milder inequality.

Our analysis implies that the existence of a monotonic
function of T,  equal to $f(T)$ at conformal fixed points,
 would lead to the inequality Eq. (\ref{eq:ineq}). 
We have already demonstrated that such a function does not exist in
general theories, for dimensions larger than two.
Even for asymptotically free theories, if such a function does exist,
it is clearly  {\em not}  $f(T)$ itself. We have been unable to find 
an alternative monotonic function for asymptotically free theories,
nor have we been able to establish its 
impossibility. Note that the existence of such a 
monotonic function, while providing a proof of Eq. (\ref{eq:ineq}), is not a 
necessary condition for the correctness of our much milder inequality.

There have been several attempts to prove a $c$-theorem in 4
dimensions
\cite{Cardy:1988cw,Jack:1990eb,Bastianelli:1996vv,Freedman:1998rd,Forte:1998dx,Anselmi:1998rd}.
The values of these $c$-functions at fixed 
points are numerically quite different from $f_{UV}$ and $f_{IR}$.
The inequalities similar to Eq. (\ref{eq:ineq}) that would arise as
consequences of these $c$-theorems in general do not significantly
constrain the 
spectrum of 4 dimensional gauge field theories. The examples of
Section 2 have already shown that our inequality {\em does} place
interesting constraints on the spectrum of 4 dimensional gauge
theories.  Other examples will be presented in Section 5.

\section{T Dependence}

We have stressed that the inequality Eq.(\ref{eq:ineq}) does not
require the monotonicity of $f(T)$, and we have noted that for one
example in which the inequality is satisfied (the supersymmetric
$SU(N)$ theory in the weak magnetic phase), monotonicity is
violated. In this section, we examine in more detail the $T$ dependence
of $f(T)$ for this example and for the other example of section two:
the non-supersymmetric $SU(N)$ theory. In each case, we record the $T$
dependence for both the $T \rightarrow \infty$ and $T \rightarrow 0$
limits, where perturbation theory may be employed.

\subsection{SUSY $SU(N)$ Theory}

For $T \rightarrow \infty$,
perturbation theory in the underlying, asymptotically free
``electric'' theory may be used, giving \cite{Grundberg:1995cu}
\begin{equation}
  \label{eq:susyTinfty}
  f(T) = f_{UV} - (N^2 - 1)(N+3F)\frac{45g_{e}^{2}(T)}{32 \pi^2}+ ...,
\end{equation}
where $f_{UV}$ is given by Eq. (\ref{eq:SQCDUV}) and where $T$ sets
the scale for the electric coupling $g_{e}$. Since the $\beta$
function is negative, $g_{e}^{2}(T)$ decreases as $T$ increases,
leading to positive $\theta$ and $\partial f
/\partial T$, as discussed in Section 3.

For $T \rightarrow 0$, with $F \le 3N/2$, perturbation theory gives
\begin{equation}
 \label{eq:susyTzero}
 f(T) = f_{IR} - ((F-N)^{2} -1)(4F-N)\frac{45g_{m}^{2}(T)}{32 \pi^2} -
3F^{2}(F-N) \frac{45y^{2}(T)}{32 \pi^2} +...,
\end{equation}
where $f_{IR}$ is given by Eq. (\ref{eq:SQCDIR}), $g_{m}$ is
the magnetic gauge coupling, and $y$ is the Yukawa coupling
of the magnetic theory. The $g_{m}^2$ term is obtained from the
$g_{e}^2$ term in Eq. (\ref{eq:susyTinfty}) by the replacement 
$N\rightarrow F-N$. The $y^2$ term is obtained by evaluating the two-loop
diagrams involving the Yukawa couplings and the
four-scalar couplings that arise from the superpotential of the magnetic
theory.

Since the theory is infrared free for $F \le 3N/2$, both couplings
increase with $T$, showing that $\theta$ and $T\partial f
/\partial T$ are negative for small $T$. Still, the inequality is satisfied.

\subsection{$SU(N)$ Theory}

We next record the $T$ dependence for the non-supersymmetric $SU(N)$
theory \cite[{\it eg}]{Smilga:1998ki}. For $T \rightarrow \infty$,
perturbation theory gives 

\begin{equation}
  \label{eq:Tinfty}
  f(T) = f_{UV} - 10 (N^2-1)(N+5F/N)\frac{g^{2}(T)}{16\pi^2}+ ...,
\end{equation}
where $f_{UV}$ is given by Eq. (\ref{eq:QCD}) and $g$ is the gauge
coupling. Asymptotic freedom leads to positive $\theta$ and
$\partial f/\partial T$.

For $T \rightarrow 0$, corrections to the free-field behavior of
the Nambu-Goldstone bosons may be computed using chiral
perturbation theory. The leading correction arises at second order
in $1/F_{\pi}^2$, where $F_{\pi}$ is the Nambu-Goldstone decay
constant, and contains a chiral logarithm. The result, for $T <<
F_{\pi}$, is \cite{Gerber:1989tt}

\begin{equation}
  \label{eq:Tzero}
f(T) = f_{IR} +
\frac{F^2(F^2-1)}{144}\frac{T^4}{F_{\pi}^4}~\ln\left(\frac{F_{\pi}}{T}\right)+\ldots, 
\end{equation}
where $f_{IR}$ is given by Eq. (\ref{eq:QCD}). Thus for small $T$,
$f(T)$ increases with $T$.

Interestingly, for the non-supersymmetric
theory in the Nambu-Goldstone phase, the function $f(T)$ is
positive-monotonic for both large $T$ and small $T$, the limits in
which it may be computed reliably, using perturbation theory.

\section{Other Examples}

In this section we apply our inequality to several example
field theories for which weakly coupled UV and IR descriptions
have been proposed. We first discuss a number of additional
asymptotically-free supersymmetric theories with infrared-free dual
descriptions. We find the inequality to be satisfied in all
cases. We then discuss the supersymmetric $SU(N)$ theory in the
regime $F > 3N/2$, where the dual magnetic theory exhibits a
non-trivial infrared fixed point. The
inequality is again satisfied where it can be checked
perturbatively. Finally, we go on to discuss QED in 2+1 dimensions,
where the inequality gives an interesting constraint on the infrared
spectrum.

\subsection{Infrared-Free Supersymmetric Examples}

All examples in this section are supersymmetric theories for
which  infrared-free dual descriptions have been proposed. We present
each example in a format where we first describe the ``electric''
theory by giving its gauge group, matter content, and superpotential.
We then give the gauge group and matter content of the dual
``magnetic'' description (and a reference to where this dual was
first described in the literature).
We show the range of flavors for which the magnetic description
exists and is infrared-free. In
each case, this is well within the flavor range for which the
electric theory is asymptotically free.
This is therefore
the regime for which free field theory calculations of $f_{UV}$
and $f_{IR}$ are exact.
We then quote the answers for $f_{UV}$ and $f_{IR}$ which are
obtained by simply counting the number of superfields and
multiplying them by a factor of $2(1+7/8)=15/4$, the contribution
to the free energy from a single free superfield.
Finally we compute the constraint following from $f_{IR} \le f_{UV}$
and check whether it is satisfied in the range of flavors for which
the calculation is valid. We find this to be the case in every
example. 

\vspace{.1in}

\noindent {\bf A.} The electric theory has $SO(N)$ gauge group with $F$
vectors and no tree level superpotential.
The magnetic dual has gauge group $SO(F-N+4)$ with $F$ vectors and
$F(F+1)/2$ meson superfields \cite{Intriligator:1995id}. As one can see
from the following table the inequality is satisfied in the entire range of
flavors where our calculation of the $f$'s is applicable. Interestingly,
as in the case of SUSY QCD the inequality is saturated at
the boundary between the conformal and free phases of the dual description,
which lies at $F=(3/2)(N-2)$.

\vspace{.1in}

\[
\begin{array}{|c|c|} \hline
\svbr {\rm range\ of\ validity}& N-2 \le F \le \frac32 (N-2) \\ \hline
\lvbr f_{UV} & \frac{15}{4} \left[ {N(N-1) \over 2} + F N \right] \\ \hline 
\lvbr f_{IR} & \frac{15}{4} \left[ {(2F-N+4)^2 \over 2} + {N-4 \over 2} \right]
\\ \hline
\svbr {\rm inequality} & F \le \frac32 (N-2) \\ \hline
\end{array}
\]

\vspace{.1in}

\noindent {\bf B.} The electric theory has $Sp(2N)$ gauge group with $2F$
fundamentals and no tree level superpotential.
The magnetic dual has gauge group $Sp(2F-2N-4)$ with $2F$ fundamentals and
$F(2F-1)$ mesons \cite{Intriligator:1995ne}. As one can see
from the table the inequality is satisfied in the entire range of
flavors where our calculation of the $f$'s is applicable.
As in the cases of $SO$ and $SU$ SUSY QCD, the inequality is saturated
at the boundary between the conformal and free phases of the dual description,
$F=(3/2)(N+1)$.
\vspace{.1in}

\[
\begin{array}{|c|c|} \hline
\svbr {\rm range\ of\ validity}& N+3 \le F \le \frac32 (N+1) \\ \hline
\lvbr f_{UV} & \frac{15}{4} \left[ N(2N+1)  + 4 F N \right] \\ \hline 
\lvbr f_{IR} & \frac{15}{4} \left[ 2(2F-N-2)^2 -N-2 \right] \\ \hline
\svbr {\rm inequality} & F \le \frac32 (N+1) \\ \hline
\end{array}
\]

\vspace{.1in}

\noindent {\bf C.} The electric theory has $SU(N)$ gauge group with $F$
flavors and an adjoint chiral superfield $A$. Without a tree level
superpotential no weakly-coupled dual is known. With the
superpotential $W=tr\ A^3$ 
a magnetic dual has been found \cite{Kutasov:1995ve} with gauge group
$SU(2F-N)$. 
The matter content of this dual is: $F$ flavors of dual quarks, a
chiral superfield 
transforming in the adjoint of the dual gauge group, and $2F^2$ mesons%
\footnote{Note that there are also known duals \cite{Kutasov:1995np}
for more general
superpotential terms $W=tr\ A^k$, but these theories do not have a weak
UV fixed point so that we cannot calculate $f_{UV}$.}. As we see
from the table the inequality is satisfied in the entire range of
flavors where our calculation of the $f$'s is applicable.

\vspace{.1in}

\[
\begin{array}{|c|c|} \hline
\svbr {\rm range\ of\ validity}& \frac{N}2 < F \le \frac23 N \\ \hline
\lvbr f_{UV} & \frac{15}{4} \left[ 2(N^2-1) + 2 F N \right] \\ \hline 
\lvbr f_{IR} & \frac{15}{4} \left[ 2(7F^2-5F N+N^2-1) \right] \\ \hline
\svbr {\rm inequality} & F \le \frac67 N \\ \hline
\end{array}
\]

\vspace{.1in}

\noindent {\bf D.} The electric theory has $SO(N)$ (or $Sp(N)$)
gauge group with $F$ vectors (fundamentals)\footnote{In the case of
$Sp$ both $N$ and $F$ are even.} and a symmetric (anti-symmetric)
tensor $T$ of the gauge group. The tree level superpotential is $W=tr\ T^3$.
The magnetic dual \cite{Intriligator:1995ff} has gauge group $SO(2F+8-N)$
($Sp(2F-8-N)$) with $F$ vectors (fundamentals), a symmetric
(anti-symmetric) tensor, and $F(F\pm 1)$ mesons.
Here and in the following the upper (lower) sign corresponds to the $SO$
($Sp$) model. Again we find that the inequality is satisfied in the
entire range of flavors where our calculation of the $f$'s is applicable.

\vspace{.1in}

\[
\begin{array}{|c|c|} \hline
\svbr {\rm range\ of\ validity}& \frac12 (N+2 \mp 8) \le F \le
\frac23 (N \mp 4) \\ \hline
\lvbr f_{UV} & \frac{15}{4} \left[ N^2-1 + F N \right] \\ \hline 
\lvbr f_{IR} & \frac{15}{4} \left[ 7F^2-5F N+N^2 \pm 41F \mp 16N+63 \right]
\\ \hline
\lvbr {\rm inequality} & N \ge {7F^2\pm41F+64 \over 6F\pm 16} \\ \hline
\end{array}
\]

\vspace{.1in}

\noindent {\bf E.} In addition to the examples above we have also applied the
predictions of the inequality to s-confining theories. These are $N=1$ SUSY
gauge theories with no tree level superpotential which confine without
chiral symmetry breaking. All s-confining theories have been
identified and their IR spectra are known \cite{Csaki:1997zb}. We find
that the confined spectra for all these theories satisfy the inequality.
Saturation occurs only for the s-confining $SU(2)$ theory with 3 flavors
which we already mentioned in the footnote of section 2.1.

\subsection{Supersymmetric Example with an Interacting Infrared Fixed Point}

We consider SUSY QCD for $F>(3/2)N$. Recall from section 2
that in this regime $f_{UV}$ is smaller than $f_{IR}$ computed
at zero (magnetic) coupling. Thus it seems that our inequality might be 
violated.
However, precisely at $F=(3/2)N$ the magnetic theory ceases to be infrared free
and instead flows to an interacting fixed point. At this 
fixed point $f_{IR}$ receives corrections from the relevant interactions. 
These corrections are calculable in perturbation theory if the fixed
point is weakly coupled and -- as we will show below-- are of the correct sign
and magnitude to ensure that the inequality holds. These results are
summarized in Figure 1. which shows
$f_{UV}$ and $f_{IR}$ as a function of $F/N$ in the neighborhood of
$F=(3/2)N$.

% FIGURE 1
\begin{figure}[ht]
  \begin{center}
    \includegraphics[width=5in]{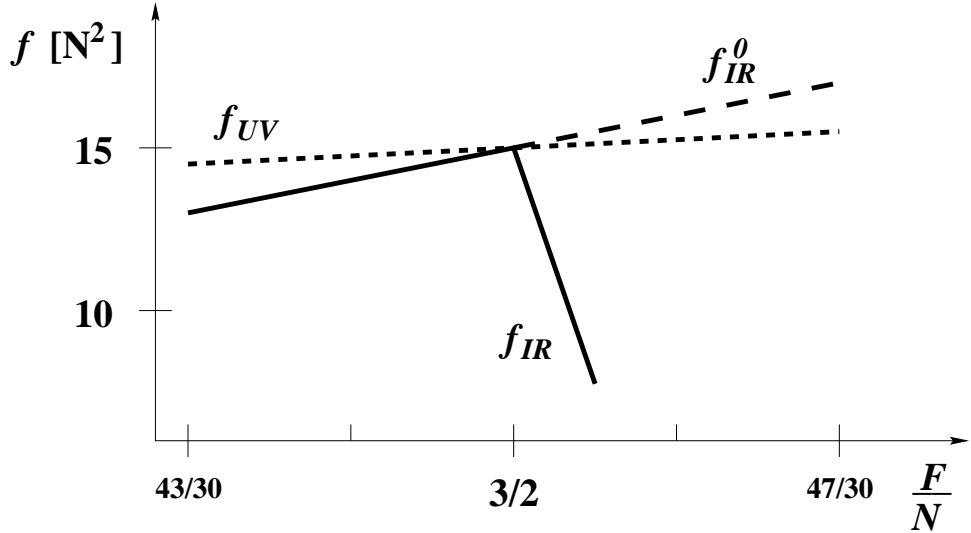}
    \caption{\it Plot of $f_{UV}$ and $f_{IR}$
      in units of $N^2$ as functions of $F/N$.
      We have taken the large-$N$ and $F$ limit and show
      only the neighborhood of the interesting point $F/N=3/2$.
      For $F/N<3/2$ one sees that $f_{IR}<f_{UV}$,
      at $F/N=3/2$ the two $f$'s touch, and for $F/N>3/2$ we again find
      $f_{IR}<f_{UV}$, but only after taking into account the interactions.
      For comparison we also show $f^0_{IR}$, the expression for $f_{IR}$
      with no interactions included.}
    \end{center}
\end{figure}

To calculate the corrections we choose large $N$ and $F$ with
$F$ tuned slightly larger than $(3/2)N$. To see that for these values of $N$
and $F$ the fixed point of the magnetic theory is
perturbative, define the small parameter $\epsilon \equiv (2F-3N)/N$ which
measures the departure (in $F$) from the free magnetic phase.
Then the fixed point values for the couplings of the magnetic
theory may be computed in terms of $\epsilon$ by setting the two-loop $\beta$
functions for the gauge and Yukawa couplings to zero. We find
\begin{eqnarray}
\label{eq:fixcouplings}
g_m^2&=& 16 \pi^2\ \frac{14}{3}\ \frac{\epsilon}{N} \nonumber \\
y^2&=& \frac{2}{7}\ g_m^2 \ ,
\end{eqnarray}
and one sees that perturbation theory in $g_m$ and $y$ holds as long as
$\epsilon \ll 1$.

We now check that the inequality is also satisfied in this interacting theory
by computing and comparing $f_{UV}$ and $f_{IR}$ at small $\epsilon$.
Equation (\ref{eq:SQCDUV}) expanded to first order in $\epsilon$ gives
\begin{equation}
f_{UV} = 15 N^2\ (1+\frac14\ \epsilon) \ .
\end{equation}
$f_{IR}$ receives contributions of order $\epsilon$ from expanding the free
theory result Eq. (\ref{eq:SQCDIR}) as well as from interactions. The 
interaction
contribution is easily obtained from Eq. (\ref{eq:susyTzero}) by
setting $g_m^2(T)$ and $y^2(T)$ equal to their fixed point values at
$T=0$, Eq. (\ref{eq:fixcouplings}). We obtain
\begin{equation}
f_{IR} = 15 N^2\ (1+ \epsilon - \frac{31}2\ \epsilon) \ ,
\end{equation}
where the $+ \epsilon$ comes from expanding the free 
expression
whereas the $-\frac{31}2\ \epsilon$ comes from the interactions. Thus
we see that our inequality $f_{UV} \ge f_{IR}$ is satisfied once
the interactions are taken into account.

A similar analysis may be used to construct a theory with a
non-trivial UV 
fixed point in which $f_{UV} < f_{IR}$\footnote{We thank Matt
  Strassler for showing us a similar theory, which led us to this
  example.}. When the number of 
flavors is just below $3 N$  the electric theory has a weakly coupled
fixed point. At this fixed point, the theory at the origin of moduli
space is in a conformally invariant phase. At this point the interactions
reduce $f$ below its free field value ({\it cf.} Eq. (\ref{eq:susyTzero})).
Away from the origin of moduli
space the UV behavior of the theory is still described by this fixed point,
while in the IR the gauge group is partially broken and some of the
flavors become massive. The infrared theory will either be free
or flow to a nonzero fixed point smaller than the UV value. 
The net difference $f_{IR} - f_{UV}$ will be positive if the number of 
flavors that have expectation values is not too large.

\subsection{QED in d=3}

For $2+1$ dimensional QED (QED$_3$) we will show that the inequality
gives an interesting constraint on the allowed infrared phase structure.
QED$_3$ with $2F$ charged Weyl fermions
($F$ Dirac fermions) is believed to have a phase transition as the number of 
flavors is varied \cite{Appelquist:1988sr,Appelquist:1995ui}. The
massless theory has a $U(2F)$ global 
symmetry. For large $F$, the screening effect of the fermions
prevents   the formation of a condensate
and the infrared theory is expected to be conformal. For small $F$,
on the other hand, one expects global symmetry breaking and
dynamical mass generation for the fermions. 
An analysis of the breaking using a gap equation indicates that 
a parity conserving mass term is formed, corresponding to the breaking of
the global $U(2F)$ symmetry to its $U(F) \times U(F)$ subgroup. 

The inequality places a tight constraint on this pattern of breaking.
QED$_3$ is free in the ultraviolet and using Eq. (\ref{eq:f}) we have
\footnote{In $(2+1)$ dimensions free bosons and fermions respectively
contribute 1 and $3/4$ to $f$.}  
\begin{equation}
f_{UV} = 1+\frac34~4F\ ,
\end{equation}
where $4F$ counts the fermionic degrees of freedom. The breaking of the 
$U(2F)$ symmetry to $U(F) \times U(F)$ leads to $2F^2$ Nambu-Goldstone bosons.
Since the theory does not confine, the photon remains in the infrared spectrum 
so we have
\begin{equation}
f_{IR} = 1+ 2F^2\ .
\end{equation}
The inequality is satisfied only for $F \le 3/2$ which implies that
chiral symmetry breaking is excluded for all $F \ge 2$.

The critical
number of flavors separating the two phases has been estimated,
using the gap equation with a $1/F$ expansion of the kernel,
to be in the range  $3 < F_{crit} <
4$ \cite{Appelquist:1988sr,Appelquist:1995ui}. The discrepancy  
between this result and our inequality suggests that 
the gap equation over-estimates $F_{crit}$.

\section{Conclusion}

We have proposed a general constraint on the structure of
asymptotically free field theories, the inequality
Eq. (\ref{eq:ineq}). Although we have  
not proven this inequality, we have shown that it agrees with a large
number of known results. In addition it places interesting
restrictions on the pattern of symmetry breaking in many cases.
The inequality (or one similar to it) would arise as a
consequence of a $c$-theorem in four dimensions, but is a weaker
condition, and can be true even in circumstances where a
$c$-theorem is not. It nevertheless  provides a constraint
on the general character of renormalization group flows for a wide
variety of asymptotically free field theories with IR fixed points.
In specific cases it may be possible to prove the inequality via the
route of section 3. 
We have noted that the inequality can be violated for 
field theories with non-trivial UV fixed points, and have provided an example 
of such a theory in Section (5.2). The inequality can also be valid
for theories with non-trivial ultraviolet fixed points, provided that
$\theta$ is sufficiently positive over a large temperature range.

Finally, it is interesting to apply the inequality to chiral gauge 
theories. In particular,  in a model due to Bars and 
Yankielowicz \cite{Bars:1981se} in which the anomaly matching
conditions are consistent  
with the formation of massless composite fermions, the inequality leads 
to a nontrivial constraint on the infrared spectrum. In a future paper 
\cite[in preparation]{CASS}, we will discuss the application of the inequality 
to this and several other chiral gauge models.

\noindent\medskip\centerline{\bf Acknowledgments}

We acknowledge the hospitality of the Aspen Center for Physics
where this work was initiated. One of us (TA) also acknowledges the
hospitality of Fermilab where he visited as a Frontier Fellow
during October and November, 1998. We thank Nima Arkani-Hamed,
Sidney Coleman, Erik D'Hoker, Dan Freedman, Erich Poppitz, Subir
Sachdev, Myckola 
Schwetz, Robert Shrock, R. Shankar, Matt Strassler and John Terning
for helpful conversations. This work was supported in part by the
Department of Energy under grant numbers DOE grant DE-FG02-91ER40676,
DE-FG02-92ER-40704 and DE-AC03-76SF00515.

%\bibliography{ct}
%\bibliographystyle{utcaps}

\providecommand{\href}[2]{#2}\begingroup\raggedright\endgroup

\end{document}